\documentclass[%
 reprint,
 amsmath,amssymb,
 aps,pra
]{revtex4-2}
\usepackage{graphicx}
\usepackage{amsmath}
\usepackage{comment}
\usepackage{siunitx}

\begin{document}
\title{An accordion superlattice for controlling atom separation in optical potentials}
\author{Simon Wili}
\author{Tilman Esslinger}
\author{Konrad Viebahn}
\email{viebahnk@phys.ethz.ch}
\affiliation{Institute for Quantum Electronics \& Quantum Center, ETH Zurich, 8093 Zurich, Switzerland
}

\begin{abstract}
We propose a method for separating trapped atoms in optical lattices by large distances.
The key idea is the cyclic transfer of atoms between two lattices of variable spacing, known as accordion lattices, each covering at least a factor of two in lattice spacing.
By coherently loading atoms between the two superimposed potentials, we can reach, in principle, arbitrarily large atom separations, while requiring only a relatively small numerical aperture.
Numerical simulations of our `accordion superlattice' show that the atoms remain localised to one lattice site throughout the separation process, even for moderate lattice depths.
In a proof-of-principle experiment we demonstrate the optical fields required for the accordion superlattice using acousto-optic deflectors.
The method can be applied to neutral-atom quantum computing with optical tweezers, as well as quantum simulation of low-entropy many-body states.
For instance, a unit-filling atomic Mott insulator can be coherently expanded by a factor of ten in order to load an optical tweezer array with very high filling.
In turn, sorted tweezer arrays can be compressed to form high-density states of ultracold atoms in optical lattices.
The method can be also be applied to biological systems where dynamical separation of particles is required.
\end{abstract}

\maketitle

\section{Introduction}
Optical lattices~\cite{gross_quantum_2017,gross_quantum_2021} and optical tweezer arrays~\cite{barredo_atom-by-atom_2016,endres_atom-by-atom_2016,ohl_de_mello_defect-free_2019,schymik_enhanced_2020,sheng_defect-free_2022,singh_dual-element_2022,jenkins_ytterbium_2022,kaufman_quantum_2021} recently emerged as prime platforms for quantum simulation and computation with neutral atoms.
On the one hand, optical lattices are generated by interference of two or more light beams, resulting in potentials defined by individual $k$-vectors.
Consequently, the resulting periodic patterns are of exceptional purity, often with negligible disorder, allowing strongly correlated phases of matter to be realised with ultracold atoms~\cite{greif_site-resolved_2016,cheuk_observation_2016,mazurenko_cold-atom_2017,hilker_revealing_2017}.
On the other hand, optical tweezers are usually projected via multiple radio-frequency (RF) tones of acousto-optic deflectors or, alternatively, from spatial light modulators.
Tweezers allow the addressing of individual atoms, including the sorting of atoms into defect-free arrays via external feedback and control~\cite{barredo_atom-by-atom_2016,endres_atom-by-atom_2016,ohl_de_mello_defect-free_2019,schymik_enhanced_2020,sheng_defect-free_2022,singh_dual-element_2022}.
When combined with Rydberg excitations, optical tweezers have been used to create synthetic many-body states~\cite{ebadi_quantum_2021,scholl_quantum_2021,browaeys_many-body_2020}.

Recent years have shown tremendous advancements in both fields, tweezers and lattices, towards the realisation of programmable and universal quantum logic~\cite{bluvstein_quantum_2022,graham_multi-qubit_2022,hartke_quantum_2022,daley_practical_2022}.
The next frontier is combining both techniques and their respective benefits -- addressability of tweezers and disorder-free, narrowly spaced potentials for lattices.
Such a combination could revolutionize neutral-atom quantum technologies and several studies have been promoting this goal~\cite{spar_realization_2022,yan_two-dimensional_2022}.
However, the fundamental mismatch in interatomic spacing between the two platforms has yet to be overcome.
While the minimum spacing between tweezers is usually on the order of a few micrometers, the spacing in optical lattices is typically several hundred nanometers.

Here, we propose a method to separate atoms in optical lattices by, in principle, arbitrary distances and we demonstrate the required optical fields to do so.
Thereby, the gap in atomic spacing between optical lattices and optical tweezer arrays can be bridged.

\section{Theoretical background}\label{sec:theory}
\begin{figure}
    \centering
    \includegraphics[width=0.35\textwidth]{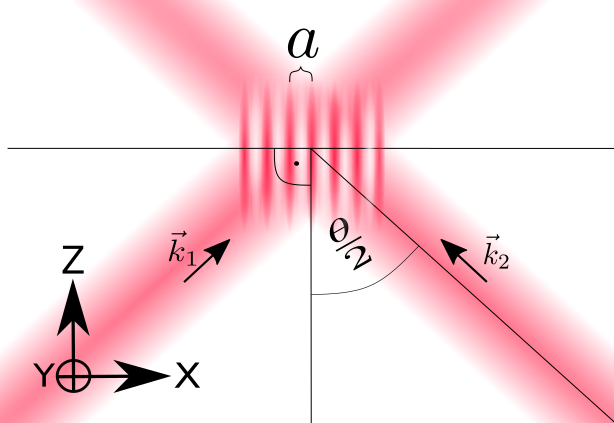}
         \caption{
An optical lattice formed by interfering two coherent beams with
        a stable phase relation. The lattice constant $a$ depends only on the optical
        wavelength and the angle enclosed by the intersecting beams. While the
        wavelength is typically fixed, the angle can be varied to create an
        optical accordion.}
         \label{fig:Beams}
\end{figure}

Optical lattices are periodic or quasiperiodic interference patterns resulting from several mutually coherent light beams.
Due to the dipole force of light on polarizable particles, such as atoms, molecules, or even macroscopic objects, the interference pattern results in a potential landscape $U$ which is proportional to the intensity of the light.
Lattices can be easiest understood by considering the interference of two plane waves, $\vec{E}(\vec{r},t)_j=\vec{E}_0e^{i(\omega t-\vec{k}_j\cdot\vec{r}+\varphi_j)}$ of equal polarization and frequency $\omega$.
The other parameters are amplitudes $\vec{E}_0$, wavevectors $\vec{k}_j$, and phases $\varphi_j$.
Let us further assume that the wavevectors $\vec{k}_j$ lie in the $xz$-plane and enclose an angle $\theta/2$ with the $z$-axis, as shown in Figure~\ref{fig:Beams}.
The resulting trapping potential reads
$$U\propto I=|\vec{E}_1+\vec{E}_2|^2\propto\cos^2\left[\pi x\left(\frac{\lambda}{2\sin\left(\frac{\theta}{2}\right)}\right)^{-1}+\Delta\varphi\right],$$
where $\lambda$ is the wavelength of the light and $\Delta\varphi\equiv\varphi_1-\varphi_2$.
A one-dimensional standing wave pattern emerges with periodicity
\begin{equation}\label{eqn:spacing}
    a=\frac{\lambda}{2}\times \frac{1}{\sin(\theta/2)}\equiv \frac{\lambda}{2}\times\frac{1}{\text{NA}}~.
\end{equation}
While the periodicity can be tuned by varying the wavelength $\lambda$ or the enclosed angle $\theta$, a change in $\Delta\varphi$ results in a translation of the lattice.
In general, a lattice with variable spacing is referred to as as an (optical) accordion lattice~\cite{williams_dynamic_2008,li_real-time_2008,al-assam_ultracold_2010,ville_loading_2017,tao_wavelength-limited_2018}.
The mathematical description of the interference pattern gets more involved when assuming realistic Gaussian beams instead of plane waves (Appendix~\ref{sec:app}).
Here we state the approximate potential landscape in the $z=0$ plane,
$$U\sim \exp\left[-2\left(\frac{\rho z_r}{w_0z_0}\right)^2\right]\cos^2\left[\pi x\left(\frac{\lambda}{2\sin\left(\frac{\theta}{2}\right)}\right)^{-1}+\Delta\varphi\right],$$
where $\rho\equiv\sqrt{x^2+y^2}$, $z_r$ is the so-called Rayleigh range, $w_0$ the beam waist, $z_0$ the distance of the beam waist to the lattice where the last three quantities are assumed equal for both beams.
Compared to the plane-wave description, the optical lattice is now confined in space by a Gaussian envelope, but the lattice spacing remains unchanged.
The one-dimensional case discussed here can easily be extended to two-- or three--dimensional lattices by adding pairs of beams with orthogonal polarization.

\begin{figure}
         \includegraphics[width=0.48\textwidth]{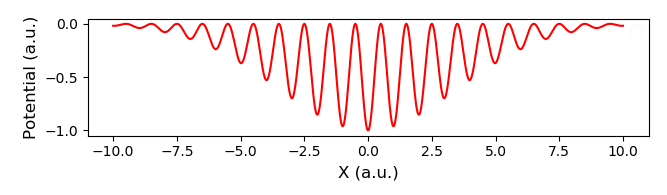}
         \label{fig:Potential}
         \caption{Transverse section through the center of a potential formed by
two Gaussian beams intersecting at an angle. We assume beams that are red-detuned with respect to the relevant optical transitions, leading to an attractive potential.
}
\end{figure}

An important figure of merit in optical accordion lattices is the tuning range of the lattice spacing and interparticle distance.
Experimentally, the tuning range is usually restricted by optical access, limiting the range of interfering angles which can be accessed.
If a single objective is used to project both lattice beams, a large numerical aperture (NA, see Eq.~\ref{eqn:spacing}) is required.
This brings about its own challenges due to optical aberrations~\cite{phelps_dipolar_2019} and limited lattice diameters.

In general, the enclosed angle can be varied by mechanically moving parts, such as translation stages~\cite{li_real-time_2008,ville_loading_2017}, rotating fiber tips~\cite{tao_wavelength-limited_2018}, or galvanometers~\cite{greif_site-resolved_2016}.
Alternatively, one can rely on acousto-optic beam steering~\cite{williams_dynamic_2008,al-assam_ultracold_2010}.
While the mechanical methods allow for a large range of angles, acousto-optic steering offers high precision but the range of achievable angles, hence lattice constants, is very limited.

In the following, we present a novel method that circumvents the challenge of limited tuning ranges of acousto-optic steering.
In addition, a scheme to avoid optical aberrations and generate large-scale lattices is developed in Section~\ref{sse:NA}.

\section{Method for achieving arbitrary large interparticle separation}\label{sec:Idea}

Let us assume an optical lattice of spacing $a_0$ uniformly filled with atoms.
We aim to stretch the lattice in space, for example to address individual lattice sites with optical tweezers or in order to overcome the diffraction-limited site-resolution of quantum gas microscopes.
Now we introduce two superimposed accordion lattices which each cover a factor of two in lattice spacing, inspired by two-frequency Floquet engineering~\cite{minguzzi_topological_2022}.
By cyclicly transferring the atoms between the two lattices and expanding one lattice at a time, we can separate the particles by, in principle, arbitrarily large distances.
\begin{figure}[b!]
\includegraphics[width=0.48\textwidth]{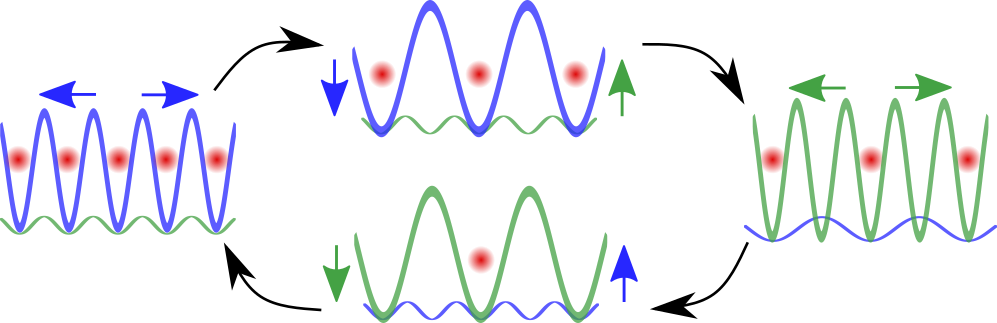}
\caption{The accordion superlattice method. It consists of two accordion lattices
operating with lattice constants between $a_0$ and $2a_0$. The atoms, molecules, or other polarizable particles, henceforth `atoms', are initially located in lattice 1 (left panel, blue lattice). This lattice is expanded by a factor of two (top).
Then the atoms are transferred into a second lattice (green) with the same initial
lattice constant $a_0$ (right).
Now the atoms occupy every second site. The second lattice can be expanded before the atoms are
loaded back into lattice 1 (bottom). Cyclic repetition of those steps allows increasing
the interatomic distance by an arbitrary factor ($>1$) utilizing two accordions
that can be expanded by a factor of two.
The two colours represent lattices whose frequencies differ by a few Megahertz, due to the diffraction of the AOD, but otherwise result from a single laser source.}
\label{fig:Idea}
\end{figure}
The scheme consists of four steps and is depicted in Figure~\ref{fig:Idea}, starting from the left-most panel.
\begin{enumerate}
    \item Stretch the initial (blue) lattice until a lattice constant of $2a_0$ is reached.
    \item Align a very shallow, second lattice with a lattice constant $a_0$ such that all the potential minima of the first lattice coincide with minima of the second lattice (green). Then ramp up the potential of the second lattice and switch off the first lattice.
    \item Stretch the second lattice by a factor of two.
    \item Transfer the atoms back to the first lattice by aligning the lattices before ramping up the first lattice while ramping down the second one.
\end{enumerate}
Those four steps can be repeated and in each cycle the distance between particles
is increased by a factor of four.
The expansion can be interrupted, for instance at step 1 or step
3, to access lattice constants between multiples of $a_0$ in the last cycle.
Thus, the interparticle distance can be
increased by an arbitrary factor larger than one.
Alternatively, the protocol can be reversed, thereby shrinking the distance between the particles.
Compared to conventional accordion schemes which aim for separations larger than a factor of two, the required numerical aperture of this method is substantially reduced.
This comes at the cost of aligning two accordion lattices precisely onto one another, which can be achieved using the precision offered by acousto-optic steering.
Moreover, the two lattices which differ by exactly a factor of two in spacing typically have a frequency difference of several Megahertz, ensuring that interferences between both lattices average out on atomic timescales.
In the following, we present an experimental setup to realise the optical fields for the accordion superlattice scheme (Sec.~\ref{sec:Setup}) while experimental measurements are shown in Section~\ref{sec:Experimental}.

\section{Setup and measurements}\label{sec:Setup}

\begin{figure}[tb]
\centering
\includegraphics[width=0.48\textwidth]{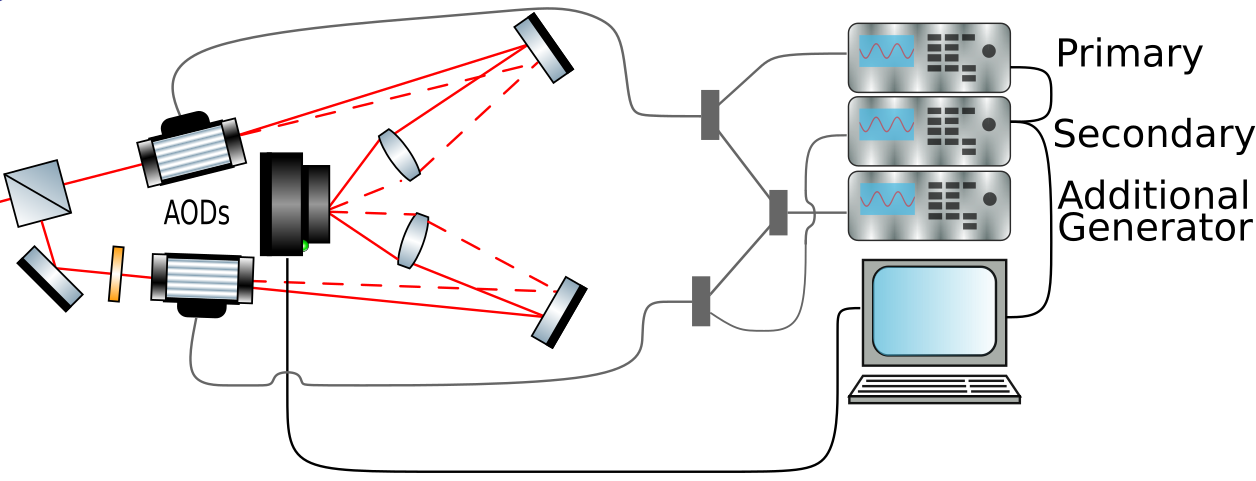}
\caption{Schematic representation of the experimental
setup. Two frequency generators are connected in a primary-secondary configuration and provide a first driving frequency for their acousto-optic modulator (AOD). An additional generator provides a second driving frequency for both AODs. The two RF tones in each AOD result in two separate beams which are illustrated by a solid and a dashed line, respectively.
The beams intersect due to two separate microscope objectives (depicted by lenses) and form optical lattices in a single plane.
A camera in the lattice plane records images which are transmitted to a computer and analysed in
real-time. The computer provides active feedback to the secondary generator.}
\label{fig:Setup}
\end{figure}

An experimental test-setup for the accordion superlattice scheme is shown in Figure~\ref{fig:Setup}.
All lattices are derived from the same laser source, which is split into two before each beam passes through its individual acousto-optic modulator (AOD).
One of the AODs is flipped with respect to the other one, such that the deflections of the first orders occur in opposite directions.
Each AOD is driven with two RF tones, resulting in two first-order deflected beams, illustrated by a solid and a dashed red line.
The two sets of accordion beams are projected onto a single plane via individual mirrors and objectives to form optical lattices.
A camera is placed in the lattice plane to record the optical potential.
The images acquired by the camera are analyzed by a computer that is used to control three frequency generators and provide feedback.
Two frequency generators are arranged in a primary-secondary configuration, each providing one driving signal for one AOD.
The third `additional' generator outputs a signal that is split and simultaneously used to drive both AODs.
The additional generator creates the first pair of deflected beams that constitutes the first lattice.
Tuning the driving frequency, the angles of deflection change, resulting in a different lattice constant.
The pair of primary and secondary signal generators serves the same purpose, outputting signals with identical frequency but controllable relative phase to their respective AOD.
This configuration of signal generators allows independent tuning the position of one accordion lattice with respect to the other.

In the following, we discuss two features of this setup which differ from typical realisations of accordion lattices.
First, due to acousto-optic deflection, rays arriving at the objectives are not parallel but skew to the optical axis.
Second, the pair of beams constituting a lattice is not passing through a single objective.
\begin{figure}[h!]
\includegraphics[width=0.48\textwidth]{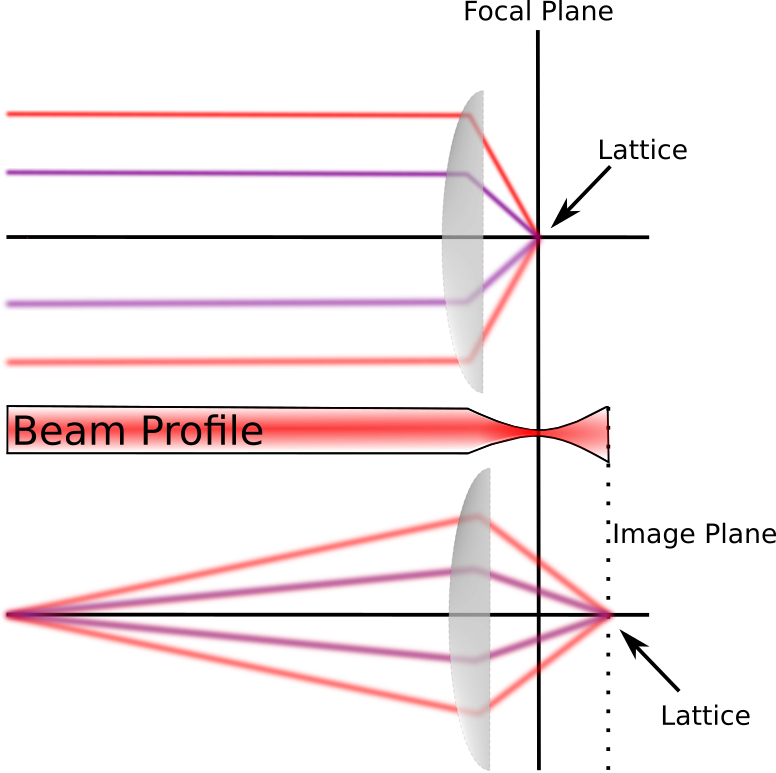}
\caption{Decoupling the image plane (IP) from the focal plane (FP). Top: conventional
schemes relying on a focusing lens use parallel incoming rays to create a
lattice in the focal plane. Those lattices often have a small diameter because a
collimated incoming beam will have its beam waist in the FP (Middle).
Bottom: our setup uses diverging beams to form the lattice in an image plane behind the FP. If the incoming beam
is collimated, the beam diameter in the IP and hence the lattice diameter can be large.
}
\label{fig:OutOfFocus}
\end{figure}

\subsection{Creating the lattice outside the focal plane}
Typical accordion lattices are created by a pair of  parallel beams~\cite{williams_dynamic_2008,li_real-time_2008,al-assam_ultracold_2010,ville_loading_2017}, approaching a lens parallel to its optical axis.
The lens then focuses the two beams in the focal plane where they interfere and form the lattice.
By varying the distance between parallel rays, the lattice constant is tuned.
Our protocol could, in principle, be realised by two sets of parallel beams. Such a configuration is shown in Figure~\ref{fig:OutOfFocus} (top panel). While this method is conceptually simple and relatively easy to align, the fact that the lattice is generated in the focal plane is often impractical.
Firstly, the diameter of the lattice is small due to the Gaussian beam profile (middle panel), assuming collimated beams at the AOD.
Secondly, the focussed light suffers from aberrations which originate from dust or scratches on the focussing lens~\cite{phelps_dipolar_2019}.
Instead, we use pairs of large, collimated beams entering the lens at an angle relative to the optical axis and the beams form the lattice in a plane which is located behind the focal plane of the lens.
Here, the beam profile is larger and the lattice can contain a large number of lattice sites.
In addition, focussed aberrations are avoided.

As mentioned above, a typical challenge of optical accordions is that a small minimal lattice constant requires a large NA of the focusing lens (Eq.~\ref{eqn:spacing}). In our configuration of skew incoming rays the image plane is further away from the lens than its focal length.
Hence, the requirement on the NA of the lens for a given minimal lattice constant is more stringent compared to the ordinary configuration.
However, the focusing task can be split between two lenses or objectives, reducing the requirements on NA, as described in the following.

\begin{figure}[tb]
\includegraphics[width=0.48\textwidth,angle=0]{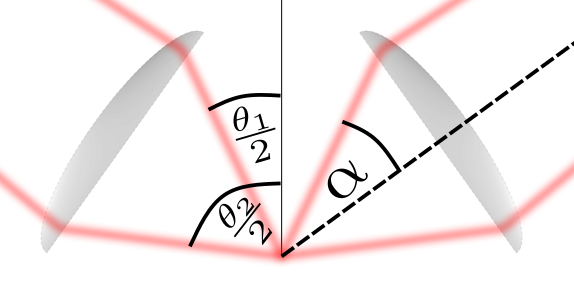}
\caption{Schematic representation of a two-lens setup. The two lower
beams will produce the smallest possible lattice constant; the two upper beams
will produce the biggest one. The numerical apertures (NA, $sin(\alpha)$), in
combination with the chosen angles ($\theta_1$,$\theta_2$), determine the range over which the
lattice constant can be tuned. Given a minimal lattice constant, using two
lenses combined with the accordion superlattice method reduces the requirement
on the NA for arbitrary atom separation substantially.
The area around the
vertical axis is kept free.}
\label{fig:SmallNa}
\end{figure}

\subsection{Lowering the requirements on the NA}\label{sse:NA}
The accordion superlattice scheme presented in Figure~\ref{fig:Idea} allows for arbitrarily large interparticle separation while requiring only relatively little optical access in the interval of angles $[\theta_1,\theta_2]$.
In practice, the achievable interparticle separation will be limited by the number of available lattice sites.
When comparing the traditional single-accordion approach to the novel accordion superlattice method, the maximal enclosed angle $\theta_2$ is the same and given by the minimal lattice constant. However, the minimal angle $\tilde{\theta}_1$ is much smaller in the single-accordion method, compared to the method employing the accordion superlattice. 
Here, the minimal required angle $\theta_1$ corresponds to a lattice constant $a=2a_0$, even for an arbitrary large interparticle separation. The angles between zero and $\theta_1$ are not used, providing optical access for other applications.
Moreover, the accordion superlattice method allows to split the projecting lens into two, as shown in Figure~\ref{fig:SmallNa}, similar to ref.~\cite{huckans_optical_2006}. This is advantageous, as each of these lenses requires only a fraction ($<\frac{1}{2}$) of the NA that a single lens would need. The requirement is initially lowered by a factor of two as two lenses are employed and then further lowered by the fact that the range of angles $[0,\theta_1]$ does not need to be covered in the accordion superlattice method. (A table of the required NAs for a few exemplary configurations is provided in the Appendix~\ref{sec:app-NA}.)

As a proof-of-principle, we use the setup described above to demonstrate the accordion superlattice scheme in the following.


\label{sec:Experimental}
\begin{figure}[t!]
\includegraphics[width=0.48\textwidth]{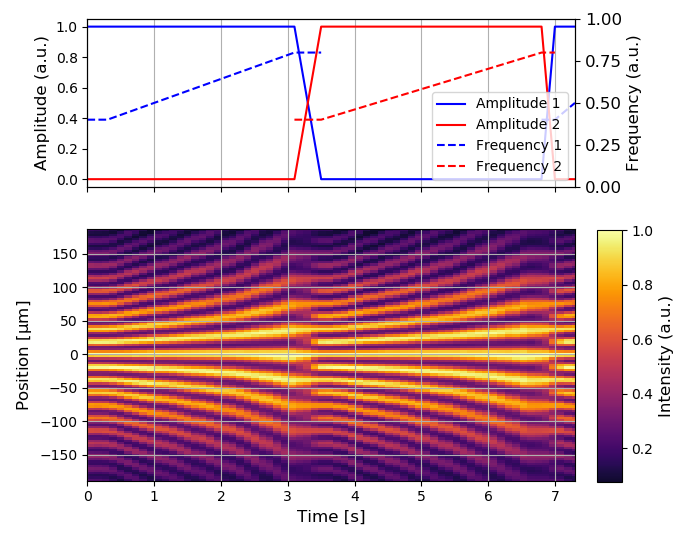}
\caption{The top panel shows the RF tones required for the
accordion superlattice method. A ramp-up of a driving frequency results in an
expansion of an accordion lattice. Increasing the amplitude of the RF tone increases the potential depth of the accordion. The initial and final
frequencies have to be tuned, such that the corresponding lattice constants
have a ratio of exactly $1:2$.
Bottom: experimental demonstration of the accordion superlattice method.
Starting from an initial lattice constant of $\SI{20.6}{\micro \meter}$, the lattice is expanded
by a factor of two within $\SI{3}{\second}$ before the second lattice is ramped up while the
expanded lattice is ramped down (between $\SI{3}{\second}$ and $\SI{3.5}{\second}$). The second lattice is then expanded before another transition takes place after $\approx \SI{7}{\second}$. The discretisation in time is due to the limited frame rate of the camera and does not reflect the expansion of the true lattice potentials, which is entirely smooth.}
\label{fig:ExperimentalDemonstration}
\end{figure}

\subsection{Experimental demonstration of the optical potential}~\label{sec:measurements}
The control protocol for the RF tones of the AODs is illustrated in the top panel of Figure \ref{fig:ExperimentalDemonstration}, realising one cycle of the hand-over process.
The measured optical potential is displayed in the lower panel of Figure~\ref{fig:ExperimentalDemonstration}. Each column of pixels, corresponding to one snap-shot, is a line-sum of the camera image. To account for the dependence of the AODs' efficiencies on the driving frequency, the measured intensities for each point in time have been normalized. 
Length and time scales in this measurement were chosen for illustrative purposes, dominated by the pixel size ($\SI{3.75}{\micro \meter}$) and the limited framerate (10 fps) of the camera.

\begin{figure}[tb]
\includegraphics[width=0.48\textwidth]{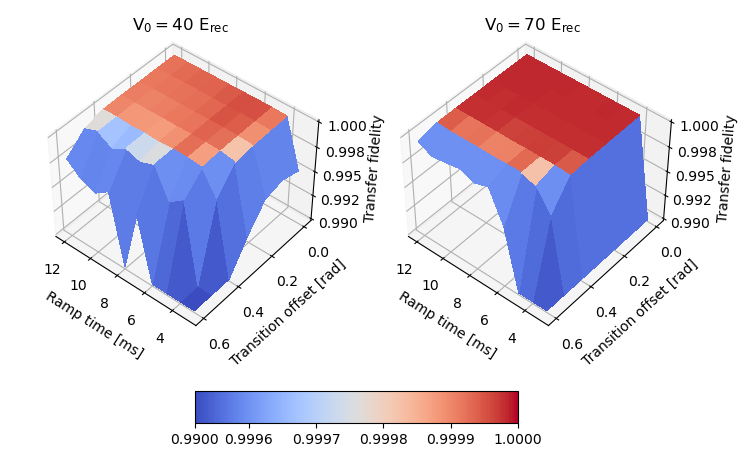}
\caption{Simulated transfer fidelity (probability of finding the atom at
the intended site after the expansion) of the accordion superlattice method. The
probability is shown for a particle at site $N=25$.
The interatomic spacing is increased by a factor of eight, i.e.~two transitions between the lattices are simulated. The transition offset is the two lattices’
misalignment during the transition process (assumed to be identical for both
transitions). On the left-hand side, the simulated initial potential depth at
the center of the lattice is 40 recoil energies; on the right-hand side, it is $70$ recoil energies.
}
\label{fig:Simulation}
\end{figure}

\section{Numerical simulation of the atomic separation}
To highlight the capabilities of the accordion superlattice scheme, we numerically simulate the transport of atoms for an expansion of the lattice by a factor of eight, that is, one and a half cycles of the hand-over scheme (Fig.~\ref{fig:Idea}).
A possible application of this process is the loading of a linear tweezer array with unit fidelity at \SI{4.2}{\micro\meter} tweezer spacing starting, for example, from a defect-free atomic Mott insulator.
The beams constituting the lattice are assumed to be Gaussian with a beam diameter of $\SI{500}{\micro \meter}$ in the lattice plane, starting with an initial lattice constant of $\SI{532}{\nano \meter}$.
The particles in the periphery of the lattice are accelerated much stronger than the inner atoms.
Moreover, the lattice depth in the outer regions is reduced due to the Gaussian profile.
Assuming a unit-filled Mott insulator across 51 lattice sites, the outermost atoms ($N = \pm 25$) are the hardest to translate.
While an atom adjacent to the trap centre only has to move by eight sites, the outermost atom effectively has to move by $8\times 25 = 150$ sites, corresponding to the extreme scenario.
Therefore, we simulate the trajectory of one (the outermost) particle, initialised as a maximally localized Wannier state of a $\mathrm{^{87}Rb}$ atom, by numerically solving the Schrödinger equation in one dimension.
We assume a lattice depth of 40 or 70 recoil energies at the central site.
The intended trajectory of the particle follows a Gaussian error function, interpolating between the initial and the final point in space.
We take the total transport time as a variable parameter.
In general, sudden acceleration or deceleration should be avoided to reduce the risk of the particle leaving its intended site.
Therefore, the transition between lattices is realised without stopping the particles. When the first lattice reaches a lattice constant of $1.8a_0$, its lattice constant is further increased to $2.2a_0$ while its amplitude is simultaneously ramped down. Meanwhile, the amplitude of the second lattice is ramped up, while it is already expanding. The alignment of the two lattices during the transition is crucial. Therefore we simulated a realistic misalignment of the two lattices during the transition, described by the `transition offset'.
A transition offset of $\pi$ rad corresponds to an offset of one full lattice site.
The misalignment is assumed to be of constant, identical absolute value but of opposite sign for both transitions.
We call the probability of finding the particle within the intended site after the expansion `transfer fidelity'.
The results of the simulation are summarized in Figure~\ref{fig:Simulation}.
We clearly see that it is possible to transfer an atom from the initial to its final intended position with high fidelity. However, it is essential that the alignment of the two lattices during transition is precise, while there is only a weak dependence on the ramp time in the investigated parameter space. For a one-dimensional lattice, we expect our simulation to be a lower bound of the individual transfer fidelity in an array of 51 atoms.
This makes the accordion superlattice scheme a promising candidate for loading tweezer arrays as large as $51\times 51= 2601$ with high fidelity.

\section{Conclusion and Outlook}
We have introduced a novel technique for tuning the distance between optically trapped particles. The method offers advantages in terms of precision, dynamic range and experimental requirements compared to known methods.
In particular, it allows arbitrarily large particle separation, limited only by optical power, while using a small numerical aperture.
Furthermore, we have demonstrated the feasibility of the optical potential in a proof-of-principle experiment and assessed the transfer fidelity with a numerical simulation.
Thus, the method can bridge the spacing gap between optical tweezers and optical lattices, allowing the transfer of atoms from lattices to tweezers and vice versa.
The setup can straightforwardly be extended to higher dimensions by applying the same concept to the orthogonal directions. In addition, mutliple RF tones can be added to superimpose more than two lattices.

The tuneable accordion lattice setup provides multiple technological applications.
First, the amplitudes and phases of both lattices with spacings $a_0$ and $2a_0$ are fully tuneable, realising a dynamic optical superlattice.
Therefore, it can be used to prepare charge-density waves~\cite{folling_direct_2007}, realise singlet-triplet oscillations~\cite{trotzky_controlling_2010,greif_short-range_2013}, and study topological pumping~\cite{lohse_thouless_2016,nakajima_topological_2016,walter_quantisation_2022}.
Second, the tuneable lattice spacing can be used to perform Bragg spectroscopy~\cite{kozuma_coherent_1999,stenger_bragg_1999,lopes_quasiparticle_2017,petter_probing_2019} and light-assisted tunnelling~\cite{aidelsburger_realization_2013,miyake_realizing_2013} over a wide range of $k$-vectors.
The frequency offset in the kilohertz-range between two lattice beams renders the diffraction angle almost unchanged, given typical centre frequencies of AODs around \SI{80}{MHz}.
Thus, both energy and momentum of the induced transitions are essentially free-tuneable.
Third, all lattice spacings between $a_0$ and $2a_0$ are readily accessible, including incommensurate frequency ratios leading to quasi-periodic~\cite{roati_anderson_2008} or quasi-crystalline~\cite{viebahn_matter-wave_2019} lattice structures.
In general, multiple RF tones can be used to compose complex optical potentials in Fourier space.
Last, the separation of atoms in small-spacing optical lattices represents a pathway towards single-site detection and addressability beyond the diffraction limit, apart from matter-wave expansion in a harmonic trap~\cite{asteria_quantum_2021}.
Applications outside quantum computing and quantum simulation are also conceivable.
For instance, the accordion superlattice method could be used to sort or assemble biological samples on different lengthscales~\cite{lee_optofluidic_2021}.

\vspace*{2ex}

\section*{Acknowledgments}
We acknowledge funding by the Swiss National Science Foundation (Grant Numbers~182650 and NCCR-QSIT) and European Research Council advanced grant TransQ (Grant No.~742579). K.V.~acknowledges financial support from the ETH fellowship.
S.W., T.E., and K.V. are authors of the patent appl.~EP22188223.6 (method and setup for changing an inter-particle distance).

\twocolumngrid


%

\clearpage

\appendix

\onecolumngrid

\section{Corrections to the lattice spacing from the interference of Gaussian beams}\label{sec:app}
\begin{figure}[h]
    \centering
    \includegraphics[width=0.45\textwidth]{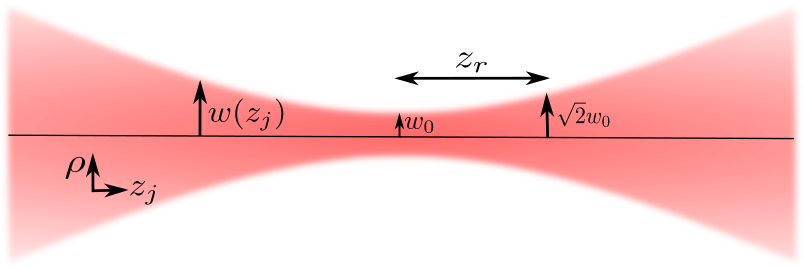}
    \caption{Transverse section of a Gaussian beam propagating along an axis
$z_j$, described by cylindrical coordinates. The minimal radius $w_0$ is called (single) beam
waist. The radius $w$ increases by a factor of $\sqrt{2}$ within one Rayleigh range
$z_r$. Far away from the beam waist, $w$ increases approximately linearly as a
function of the distance to the beam waist}
    \label{fig:GaussianBeam}
\end{figure}

Here, we provide the mathematical derivation of the potential landscape corresponding to an optical lattice formed by two interfering Gaussian beams.
Each of the Gaussian beams is conveniently described in cylindrical coordinates; we denote the radial coordinate of beam $j\in{1,2}$ with $\rho_j$, and the axial coordinate with $z_j$.
The electric field reads 
\begin{eqnarray}
\nonumber        E_j(\rho_j,z_j,t)&=&
        \hat{\varepsilon}A_{0}\frac{w_0}{w(z_j)} e^{-(\rho_j/w(z_j))^2} e^{ik\rho_j^2/2R(z_j)} \times e^{i(k_jz_j- \eta(z_j)+\varphi_j)}\\
\nonumber        \text{with}~~z_r&=&\frac{\pi w_0^2}{\lambda}\\
\nonumber        w(z)&=&w_0\sqrt{1+\left(\frac{z}{z_r}\right)^2}\\
\nonumber        R(z)&=&z(1+(z_r/z)^2)\\
\nonumber        \eta(z)&=&\tan^{-1}(z/z_r),
\end{eqnarray}
where $\hat{\varepsilon}$ is called polarization vector, $A_0$ denotes the amplitude, $k_j$ the wavewector, and $\varphi_j$ the phase offset. These quantities are also used to describe plane waves. In addition, we
have the following new terms: the beam radius $w$, its single-beam waist $w_0$, the Rayleigh range $z_r$, the radius of curvature $R$, and the Gouy phase $\eta$. A
longitudinal section of such a beam is depicted in Figure~\ref{fig:GaussianBeam}.

If the Gaussian beam has a large beam waist ($w_0\gg \lambda$), then its Rayleigh range is much longer than its wavelength ($z_r\gg \lambda$), i.e.~the intensity depends only weakly on $z_j$ in the beam
waist’s vicinity. Moreover, we have $\eta \approx 0$ and $R\rightarrow \infty$. Consequently, we
essentially have plane waves with a Gaussian envelope in radial direction. However, the situation becomes more complicated when we let the beams interfere away from their beam waists where their wavefronts are curved. This is the case for the setup described in Section \ref{sec:Setup}. For the sake of simplicity, we will restrict
ourselves to the case that was implemented experimentally. Therefore, we assume that the beam waists of both beams lie an equal distance of $z_0$ in front
of the lattice. Let us further assume that the centers of both beams cross at the
origin of a Cartesian coordinate system $x = y = z = 0$, and $z_1 = z_2 = z_0 > 0$
in the cylindrical coordinate systems of both beams. Additionally, we assume
that
$z_r \ll z_0$ and
$x,y,\rho \ll z_0$.
Those conditions can be described as having the waist far away from the lattice compared to the Rayleigh range and the lattice’s diameter being small compared to its distance to the beam waist. The lattice is created in the $xy$-plane, where the following relations hold:
\begin{equation}
    \begin{split}
        z_{1,2}(x,y)&=z_{1,2}(x)=z_0\pm \sin\left(\frac{\theta}{2}\right)x\\
        \rho_{1,2}(x,y)&=\sqrt{\left[x\cos\left(\frac{\theta}{2}\right)\right]^2+y^2}\\
        w(z)&\approx w_0\frac{z}{z_r}\\
        R(z)&\approx z\\
        \eta(z)&\approx \rm const.
    \end{split}
\end{equation}

In this regime, we find 
\begin{eqnarray}
\nonumber I(\rho,z_j=z_0)&=&I(x,y)=|E_1(x,y)+E_2(x,y)|^2\\
\nonumber&\approx&\underbrace{|A_0|^2\left(\frac{z_r}{z_0}\right)^2}_{:=I_0/4}e^{-2\left(\frac{\rho z_r}{w_0z_0}\right)^2}\cdot|e^{ik\left(\frac{\rho^2}{2z_1}+z_1\right)}+e^{ik\left(\frac{\rho^2}{2z_2}+z_2\right)}|^2\\
\nonumber&=& I_0/4\cdot e^{-2\left(\frac{\rho z_r}{w_0z_0}\right)^2}\left(2+2\cos\left[k\{\frac{\rho^2}{2}(\frac{1}{z_1}-\frac{1}{z_2})+z_1-z_2\}\right]\right)\\
\nonumber&=& I_0e^{-2\left(\frac{\rho z_r}{w_0z_0}\right)^2}\cos^2\left[\frac{k}{2} \{\frac{\rho^2}{2}(\frac{1}{z_1}-\frac{1}{z_2})+z_1-z_2\}\right]\\
\nonumber&=& I_0e^{-2\left(\frac{\rho z_r}{w_0z_0}\right)^2}\cos^2\left[\frac{k}{2} \{\frac{-\rho^2\sin(\theta/2)x}{z_0^2-\sin^2(\theta/2)x^2}+2\sin(\theta/2)x\}\right]\\
 &=& I_0e^{-2\left(\frac{\rho z_r}{w_0z_0}\right)^2}\cos^2\left[\frac{\pi}{\underbrace{\left(\frac{\lambda}{2\sin(\theta/2)}\right)}}_{=a}x \left(1-\underbrace{\frac{1}{2}\frac{\rho^2}{z^2_0-\sin^2(\theta/2)x^2}}_{\rm correction\: term}\right)\right].
\end{eqnarray}

Note that the approximation we made in the second line neglects minor effects regarding the contrast, but the value of the lattice constant is exact. In the last line, we find the familiar form of a $\cos^2$ potential, up to a correction term. The correction term makes the lattice constant larger towards the edge of the lattice. Its effect is tolerable, as we show in the following. During the transition, at a position $x,y$, the two lattices have a phase offset due to the correction term given by 
\begin{eqnarray}
\nonumber
|\delta\varphi| &=&\left| \frac{\pi}{\left(\frac{\lambda}{2\sin(\theta/2)}\right)}x\frac{1}{2}\rho^2\left(\frac{1}{z_0^2-\sin^2(\theta/2)x^2}-\frac{1}{2z_0^2-\frac{1}{2}\sin^2(\theta/2)x^2}\right)\right|\\
&\approx & \left| \frac{\pi}{4\left(\frac{\lambda}{2\sin(\theta/2)}\right)}\frac{x\rho^2}{z_0^2}\right|
\end{eqnarray}
where $\theta$ denotes the larger of the two enclosed angles.
For realistic values of $x/a=100$ and $\rho/z_0=0.035$ during the last lattice transition of the outermost atom of interest, we find $|\delta\varphi|<0.1$, which according to our simulation is well within the range of tolerance. To further mitigate this effect, a compensation term linear in $x$ (note that $\rho^2=x^2+y^2$) can be introduced by choosing the lattice constant of the second lattice slightly larger than two times the lattice constant of the first lattice. While this results in an increase of said offset for the atoms close to the center of the lattice, the maximal phase offset experienced by the atoms that are affected the most can be reduced by a factor larger than two.

In conclusion, the plane-wave lattice with a Gaussian envelope is a good model for the lattice created at the beam waists of two Gaussian beams. Far away from the beam waists, this still holds true approximately if the diameter of the lattice is small compared to its distance to the beam waist.
\newpage
\section{Exemplary NA requirements}\label{sec:app-NA}
\setlength{\tabcolsep}{15pt}
\begin{table}[h!]
\centering
\begin{tabular}{c c c }
\hline
\hline
$a_0/\lambda$ & naive NA & accordion superlattice NAs \\
\hline
$0.5$ & $1 $ & $0.5$ \\
$0.75$ & $0.67$& $0.19$\\
$1$ & $0.5$ & $0.135$ \\
$1.5$ & $0.33$& $0.086$\\
$2$ & $0.25$ & $0.064$ \\
\hline
\end{tabular}
\caption{The table shows the required NAs for an increase in interparticle distance by an arbitrarily large factor, starting from an initial spacing $a_0$. The middle column states the required NA when relying on a single lattice projected by a single objective. The right column contains the required NAs when using the accordion superlattice method, relying on two objectives.}
\label{table:ThetaDynamicRange}
\end{table}

\end{document}